\documentstyle[epsf]{elsart}

\begin{document}
\begin{frontmatter}

\title{J/$\psi$ Suppression in Heavy Ion Collisions\\ at the CERN SPS}  
\author[SUNYSB]{D.~E.~Kahana} \and \author[BNL]{S.~H.~Kahana}
\address[SUNYSB]{Physics Department, State University of New York at Stony Brook,
Stony Brook, NY 11791, USA}
\address[BNL]{Physics Department, Brookhaven National Laboratory,
Upton, NY 11973, USA}

\begin{abstract}

We reexamine the production of J$/\psi$ and other charmonium states for a
variety of target-projectile choices at the SPS, in particular for the
interesting comparison between S+U at $200$ GeV/c and Pb+Pb at $158$ GeV/c as
observed in the experiments NA38 and NA50 respectively. For this study we use
a newly constructed cascade code LUCIFER II, which yields acceptable
descriptions of both hard and soft processes, specifically Drell-Yan and
meson production. This code divides the ion-ion collision into an initial
phase involving hard interactions of the original nucleons and no soft energy
loss, followed after the meson formation time by a `normal' low energy
cascade among the secondary particles.  The modeling of the charmonium states
differs from that of earlier workers in its unified treatment of the hidden
charm meson spectrum, which is introduced from the outset as a set of coupled
states $\left\lbrace\psi, \chi^{i}, \psi'\right\rbrace$.  The result is a
description of the NA38 and NA50 data in terms of a conventional, hadronic
picture.  The apparently anomalous suppression found in the most massive
Pb+Pb system arises in the present simulation from three sources: destruction
in the initial nucleon-nucleon cascade phase, use of coupled channels to
exploit the larger breakup in the less bound $\chi^{i}$ and $\psi'$ states,
and comover interaction in the final low energy phase. 

\end{abstract}

\end{frontmatter}

\section{Introduction}  
The possible use of J$/\psi$ suppression as a signal of unusual behaviour in
relativistic ion collisions, first suggested by Matsui and Satz
\cite{Matsui}, has attracted considerable experimental and theoretical study.
Great interest has attached to the results obtained by the NA50 collaboration
for charmonium production in Pb+Pb collisions at $158$ GeV/c: to the early
findings presented at the Quark Matter 1996 meeting \cite{NA50a} as well as
to the startling data later released at RHIC'97 \cite{NA50b}.  The success of
Glauber-like calculations of J/$\psi$ production and breakup in the p+A and
S+U \cite{Huefner,Kharzeev,Gavin} systems, coupled with a failure of Glauber to
provide an equally good description of the apparently accelerated absorption
in Pb+Pb has been widely interpreted \cite{NA50a,NA50b,Kharzeev} as a signal
of QCD plasma creation in these collisions. The very sharp behaviour of the
J/$\psi$ yield as a function of transverse energy $E_t$ seen in the later
experiment \cite{NA50b} has especially attracted attention. 

\begin{figure}
\vbox{\hbox to\hsize{\hfil
\epsfxsize=5.5truein\epsffile[24 85 577 736]{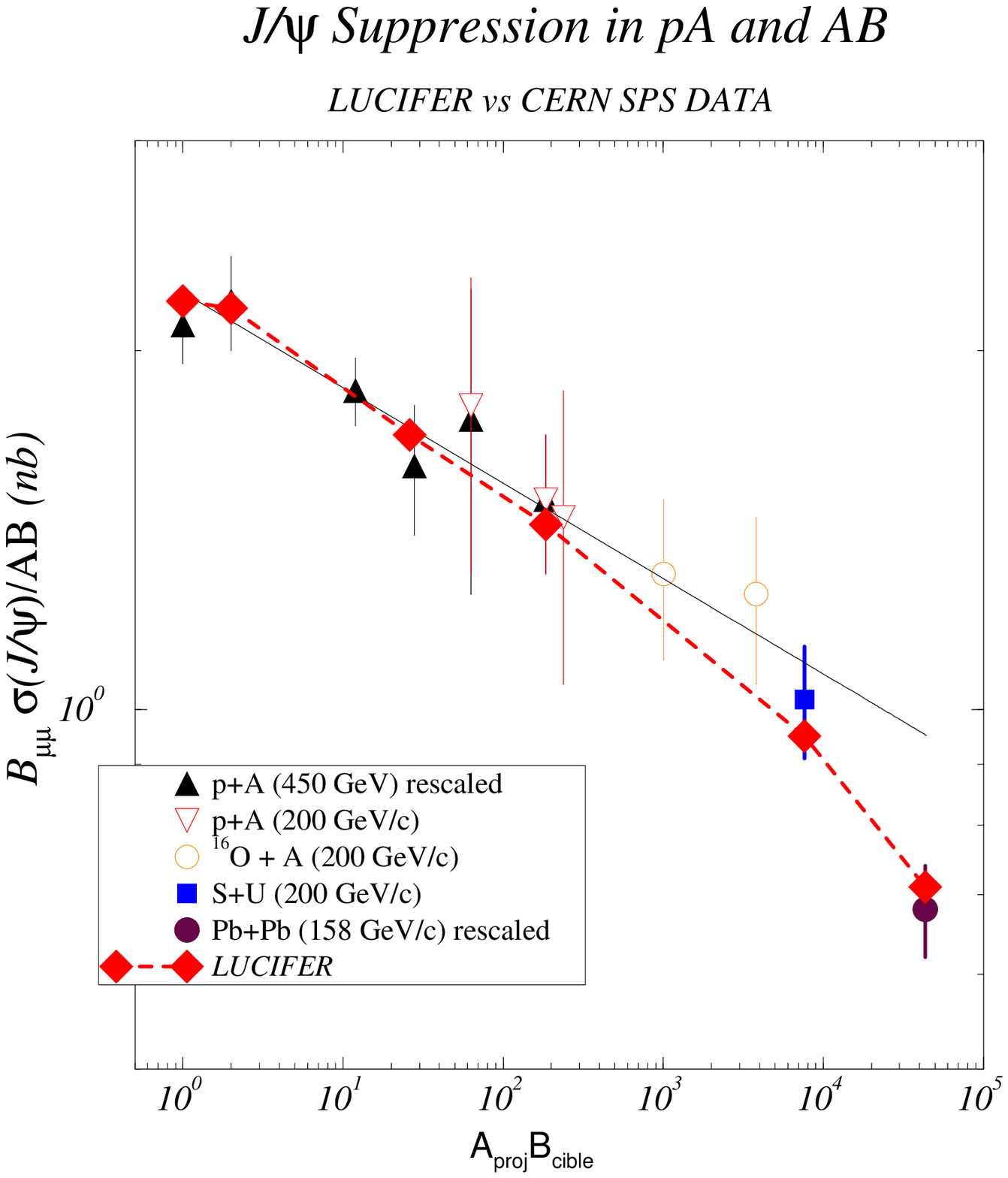}
\hfil}}
\caption[]{The whole range of yields for J/$\psi$ from $pp$ and p+D to 
Pb+Pb calculated in the cascade and compared to SPS measurements at various
energies. The absolute theoretical values are obtained by normalisation to 
nucleon-nucleon.}
\label{fig:jpsi-minbias}
\end{figure}

\begin{figure}
\vbox{\hbox to\hsize{\hfil
\epsfxsize=6.4truein\epsffile[24 85 577 736]{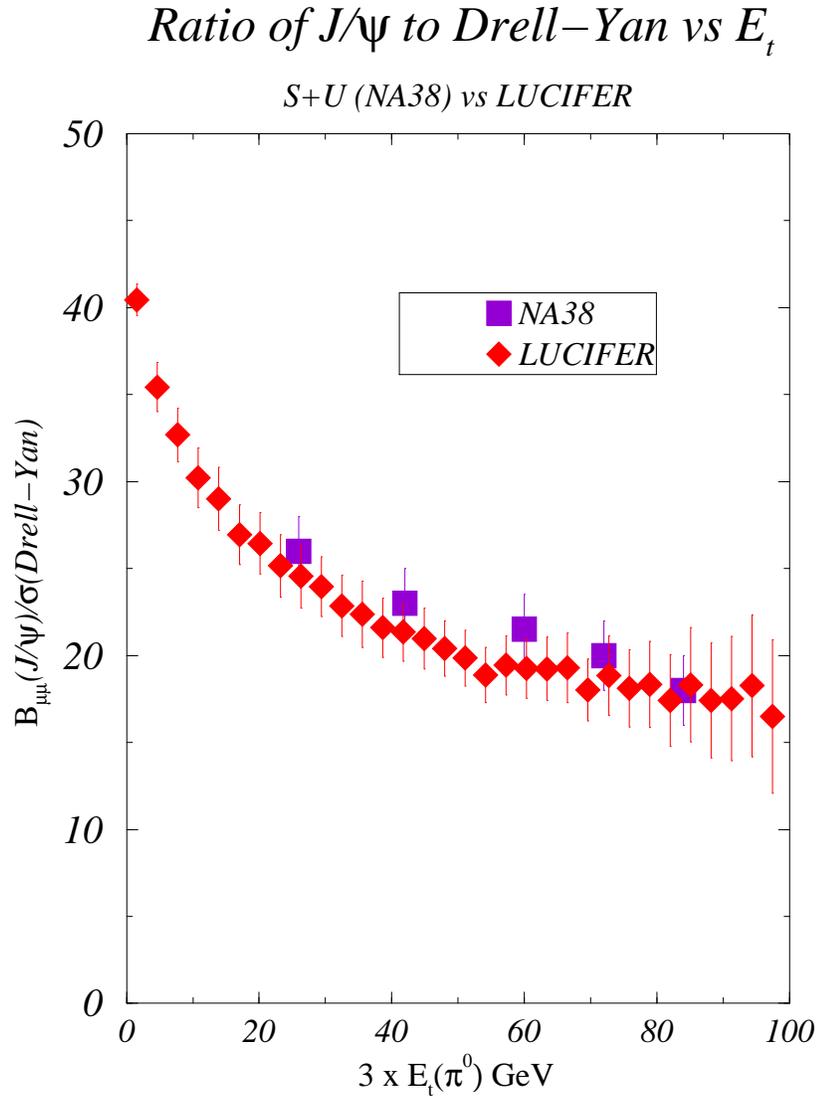}
\hfil}}
\caption[]{Comparison between the cascade and NA38 transverse energy 
dependence for J/$\psi$, for S+U.}
\label{fig:SUcentral}
\end{figure}

\begin{figure}
\vbox{\hbox to\hsize{\hfil
\epsfxsize=6.4truein\epsffile[24 85 577 736]{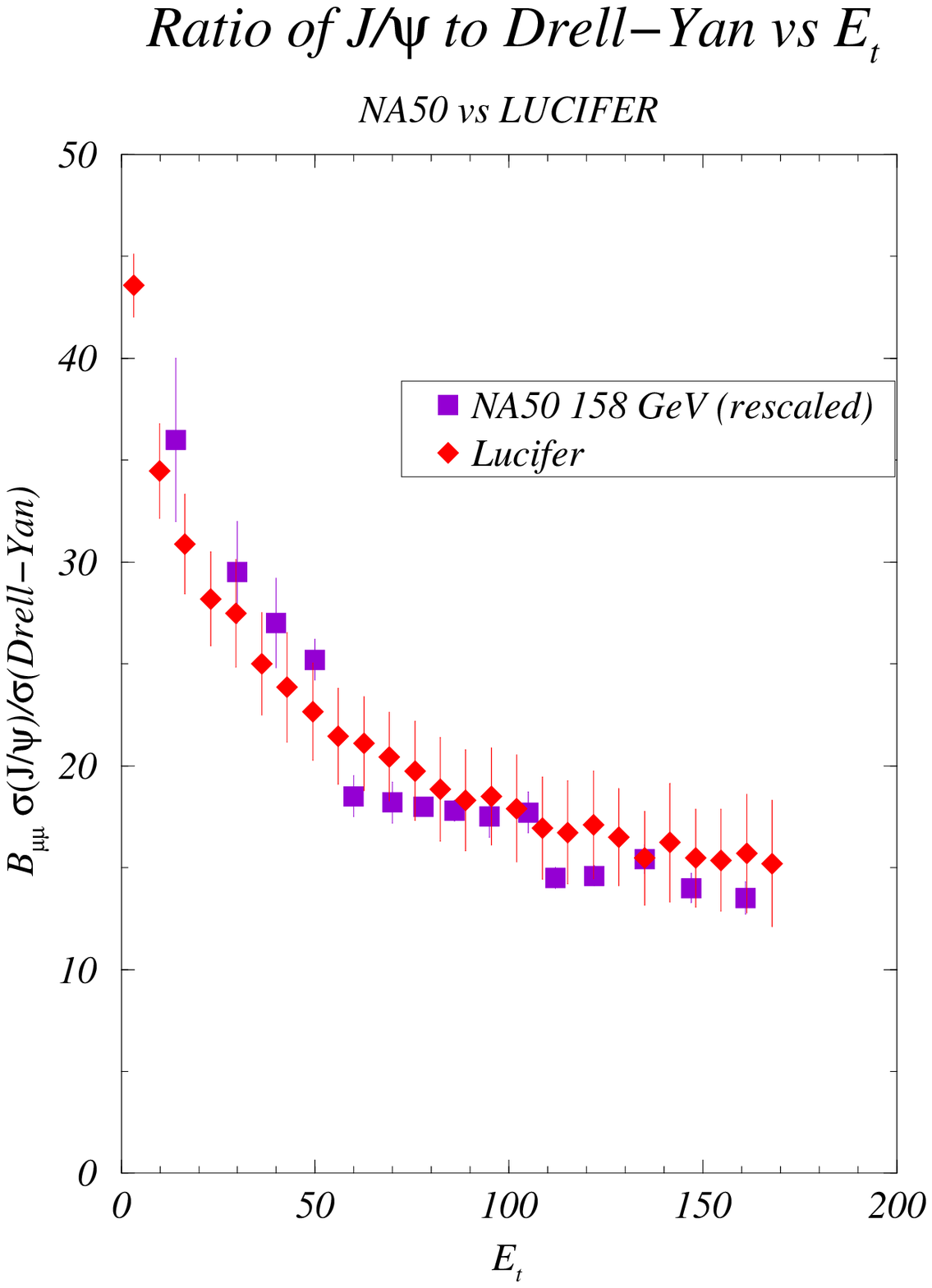}
\hfil}}
\caption[]{Comparison between the cascade and NA50 transverse energy 
dependence for J/$\psi$. There are no discontinuities, of course, in the
LUCIFER yields, but the general shape is reproduced. In both this plot and
that for central S+U, the $E_t$ scale is that obtained directly from the
cascade, no scale change has been made.}
\label{fig:PbPbcentral}
\end{figure}

We attempt to retrace this ground theoretically, employing a new, two phase
cascade approach, described in detail elsewhere \cite{LUCIFERI,LUCIFERII},
combined with a variation of the Satz-Kharzeev model for production and
annihilation of charmonium in the initial baryonic collisions. This modeling
described below, allows the coupled-channel aspect of the hidden charm
spectroscopy, $\left\lbrace \psi, \chi^{i}, \psi'\right\rbrace$ to play a
more central role. The comparison with Glauber theory based models is done with
two purposes in mind: first to understand the differences with the cascade if
any, and second to help place the cascade on a firmer foundation,
paradoxically, by indicating how similar the first high energy phase of the
cascade is to the Glauber model. In this first application, we include
partons in a minimal fashion, to describe for example Drell-Yan
production. Hence we are testing a `purely' hadronic description of the
anomalous Pb+Pb measurements. From the evidence presented in 
figures (\ref{fig:jpsi-minbias}), (\ref{fig:SUcentral}),
(\ref{fig:PbPbcentral}), it would appear such a test is justified.

It has been pointed out that a hadronic picture might succeed \cite{Gavin}
without invoking quark-gluon plasma (QGP) creation, if at least part of the
seemingly anomalous suppression in Pb+Pb could be produced by comover
annihilation, {\it i.e.\ }by interactions of the J/$\psi$ with secondary
mesons generated in the ion-ion collision. The second phase in LUCIFER II,
which is a low energy cascade, perforce includes the effect of J/$\psi$
destruction through such comover rescattering.

We begin with a description of the motivation behind LUCIFER II and a brief
outline of the two step cascade. We attempt to separate hard and soft
processes by time scale, so as to permit partonic and hadronic cascading to
be joined naturally, in a modular fashion. The separation is effected through
the use of a short time scale, automatically present at high energies: the
time $T_{AB}$ taken for the two interacting nuclei A and B to traverse each
other in the global collision frame. The uncertainty principle allows hard
interactions involving sufficiently high energy-momentum transfer, {\it i.e.\
} for $Q^{-1} \le T_{AB}$, to take place in the first and very rapid
cascade. Soft processes involving low tranverse momentum are not completed
until later. Thus in the initial fast cascading the nucleons {\it lose no
energy} but are still aware of the number and nature of the two-particle
collisions they have undergone.

Specifically, the method \cite{LUCIFERII} consists of running the cascade in
two stages. The first is a high energy fast-time mode in which collision
histories are recorded and fast processes (Drell-Yan and charmonium
production) are allowed to occur. Using the entire space-time and
energy-momentum history of this stage, a reinitialisation of the cascade is
performed using elementary hadron-hadron data as a strict guide. The final
positions and momenta of baryons in the first phase, and the number of
collisions they suffer are recorded and used to generate produced mesons
together with their initial momentum and space-time coordinates. In the
initial ion-ion collision the interacting nucleon paths are almost along
light-cones. The second cascade begins at $T_{AB}$, the time of the last
nucleon-nucleon collision, with initial conditions specified by the
reinitialisation, but no secondary interactions are allowed until a formation
time for produced mesons has passed.  The participants in the second phase
are generic mesons, thought of as of $q\bar q$-like in character with masses
centered near $M_{q\bar q}=700$ MeV and in the range $M_{q\bar q} \sim
0.3-1.0$ GeV. Generic baryons consisting of $qqq$ are also included and are
excited to rather light masses, $M_{qqq}\sim 0.94-2.0$ GeV \cite{LUCIFERII}.
All the generic hadrons decay {\it via} sequential pion emission.  Normal
stable mesons and baryons are also present, and terminate the decay chains.

Many cascades \cite{frithjof,werner,wang,genericparton,RQMD,URQMD,ARC1,LUCIFERI} 
have been constructed to consider relativistic heavy ion collisions. Since
the eventual aim of experiments designed to study such collisions is the
creation of a regime in which the quark-gluon structure of hadronic matter
becomes evident, it will ultimately be necessary to include the partonic
degrees of freedom in such cascades. However, since at SPS and even at RHIC
energies it is by no means clear that all initial or subsequent hadron-hadron
collisions occur with sufficient transverse momentum to free all partons
\cite{eskola}, at least a part of the eventual simulation must deal with
collisions both of the initially present baryons, nucleons in fact, and
finally of all the produced mesons as well.

Kharzeev and Satz \cite{Kharzeev} employ a model based on hadronic Glauber
theory describing production and breakup of the J/$\psi$ in ion-ion
collisions, to demonstrate that such a picture cannot account for the degree
of suppression seen in Pb+Pb collisions at the SPS. Reasoning similarly, we
can make a close comparison of our treatment with their work. The required
initial production of a $c \bar c$ pair is handled within an effective
hadronic formulation both in our work and in that of Kharzeev, {\it et.al.\
}. There are, naturally, specific and important differences between Glauber
theory and a cascade model, and it is partly these differences which permit
the so-called anomalous suppression in Pb+Pb to be explained within a purely
hadronic framework. Additionally, the cascade provides a real calculation of
$E_t$ as a function of centrality, {\it i.e.\ }impact parameter, with no
adjustable parameters available for producing agreement with the experimental
measurements.  Equally important, the interaction of $c\bar c$ states with
comoving secondary mesons is treated dynamically in the soft phase of our two
step cascade.

The overall degree of suppression in Pb+Pb, insofar as it differs from
earlier work \cite{Kharzeev,Gavin}, results from a combination of effects;
these are baryonic, coupled channel and comover in kind, with substantial
contributions arising from both phases of the cascade. There are potential
unknown variables: the production and dynamic time evolution of each
charmonium state, the breakup probabilities against both baryons and mesons,
the density of secondary mesons. This last is to a large extent predicted by
the cascade, which must agree with actual inclusive final state meson and
baryon distributions.  There also exist constraints on the basic charmonium
variables. The production is in principle determined in elementary
nucleon-nucleon collisions, the baryonic breakup in p+A collisions. The
$\psi \pi$ breakup cross-sections are not known directly from any
measurements. If the relative $\psi$ production in $NN$ and
$\pi N$ systems may be taken as a guide here, the $\psi \pi$ breakup
cross-sections could be expected to be directly comparable. We have used 
a factor $2/3$ to relate the charmonium-meson to charmonium-baryon breakup 
cross-sections, but also employed equal charmonium-meson as a test.

The success obtained in describing the meson spectra has already been
presented in \cite{LUCIFERII}, and of course is relevant to the degree of
charmonium destruction by comovers. In particular, the selection of a meson
formation time is tied to this latter issue. The differences between S+U and
Pb+Pb, which exhibit a considerable increase in the product $A\times B$, arise from
both baryonic processes and meson production.

The source code to LUCIFER II is made available on the world wide web at
{\sl http://bnlnth.phy.bnl.gov}, and may be downloaded either directly from
the web site or by anonymous ftp to {\sl bnlnth.phy.bnl.gov}. It is in C and
should be relatively easy to port to any UNIX system. Linux, AIX, SunOS, IRIX
and HP/UX ports have at one time or another been made.

\section{The Two Phase Cascade} 

\begin{figure}
\vbox{\hbox to\hsize{\hfil
\epsfxsize=5.0truein\epsffile[24 85 577 736]{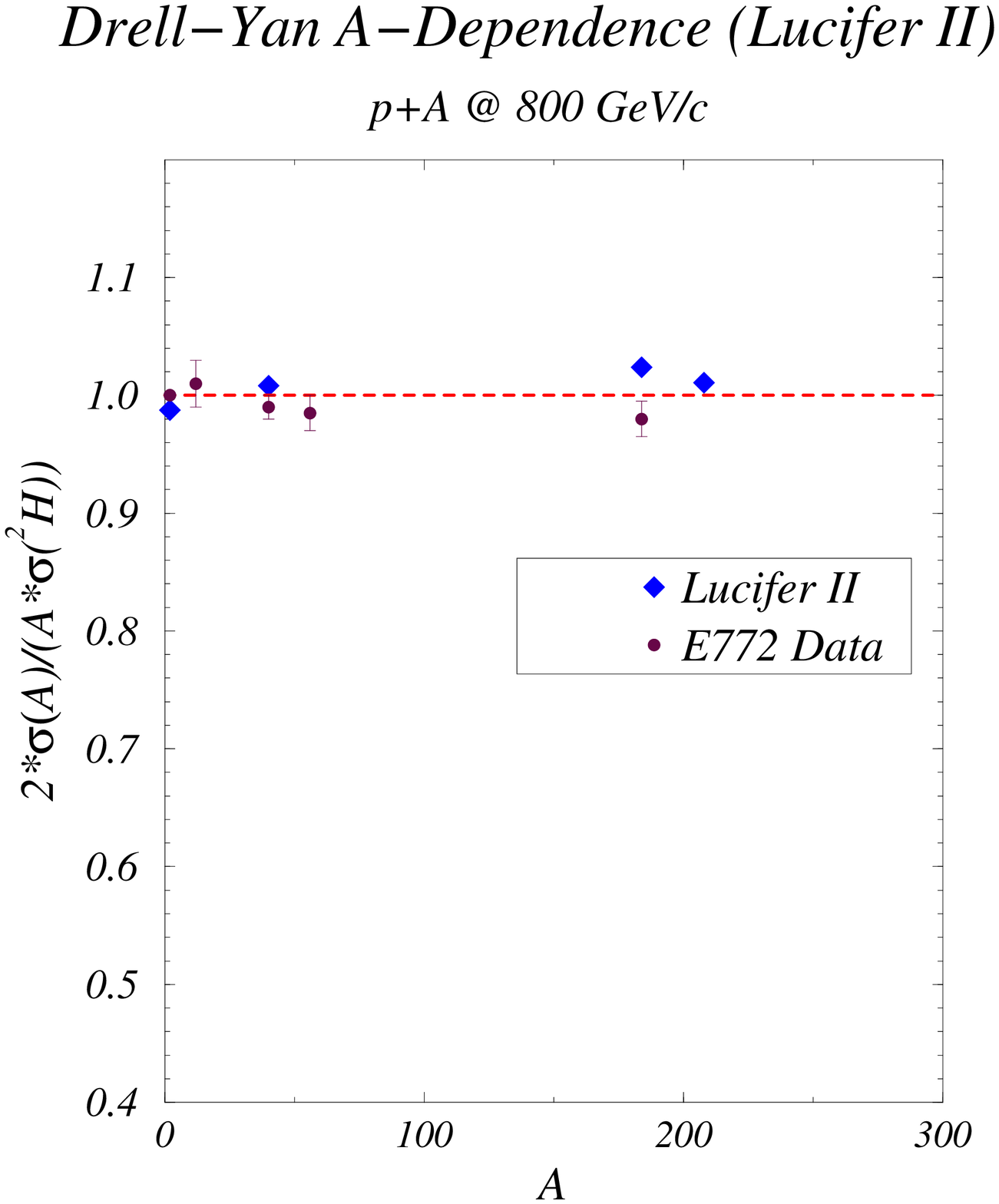}
\hfil}}
\caption[]{A-dependence of Drell-Yan at 800 GeV/c: E772 (FNAL) vs LUCIFER.
Minimum bias dimuon production as shown is calculated microscopically
using NA3 structure functions \cite{NA3} but could as well have been obtained
directly from considerations of the total collision number. The same argument
applies to the high mass dimuon cross-section as a function of $E_t$.}
\label{fig:one}
\end{figure}

We present in this work a mere outline of the cascade architecture, details
having been provided in earlier work \cite{LUCIFERII}.  We already noted the
global time scales which divide the cascade into two steps, the first loosely
designated as `hard', the second as `soft'.  This separation has been
discussed often in the literature
\cite{gottfried,koplik,amueller,bjorken,dokshitzer}.  Energy loss and meson
production associated with low transverse momentum $p_t$ are slow, `soft'
processes.  In contrast stand fast or `hard' processes, involving large
$p_t$, of which production of high mass Drell-Yan pairs \cite{NA3,E772,NA51}
is a good example. The $A$-dependence of minimum bias Drell-Yan data
\cite{E772} in p+A collisions, see Fig(\ref{fig:one}), suggests that high
mass lepton pair production occurs only at the highest collision energy. For
a theory of charmonium suppression to be taken seriously, it must
simultaneously explain this striking feature of Drell-Yan data, and the
substantial soft energy loss experienced by the projectile nucleon.  If not,
an $A$-dependent suppression in p+A might be built into the model,
attributable to production in successive projectile-target nucleon-nucleon
collisions occurring at lower and lower energies, where the chance of J/$\psi$
production is considerably less. The present model does well in this regard
as seen in figure \ref{fig:one}.

\begin{figure}
\vbox{\hbox to\hsize{\hfil
\epsfxsize=5.0truein\epsffile[24 59 562 736]{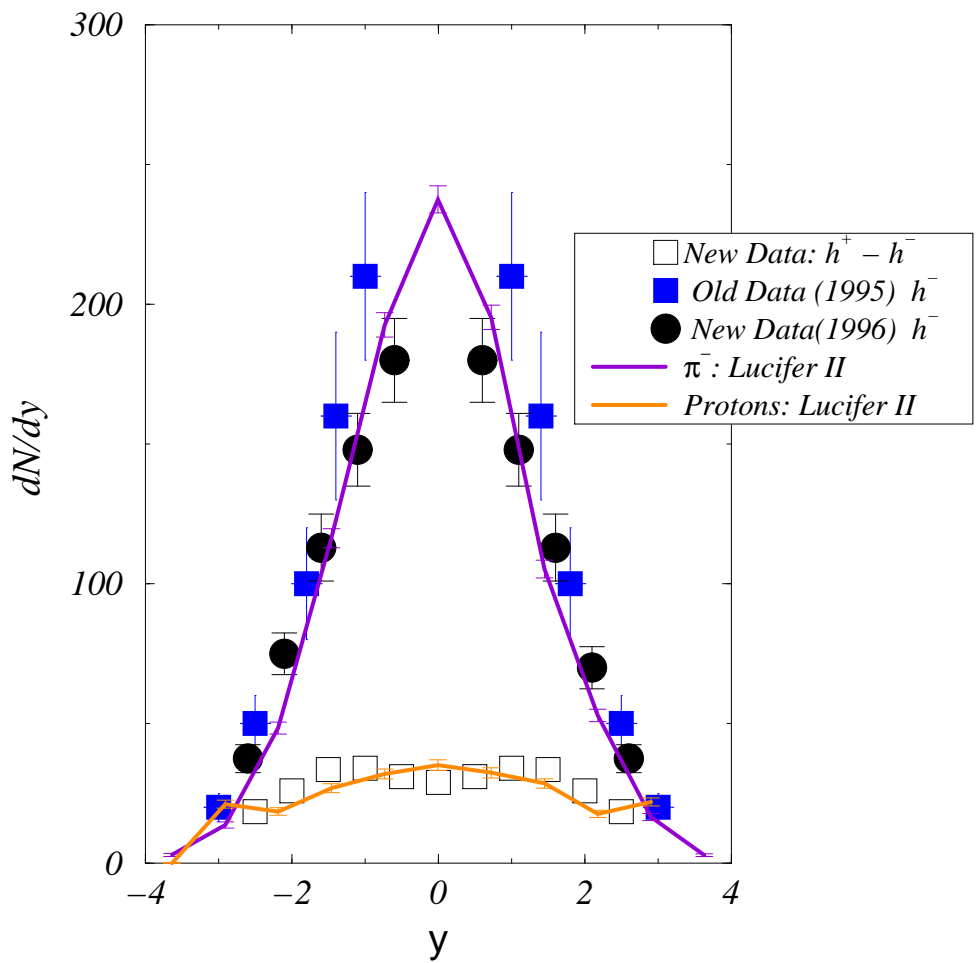}
\hfil}}
\caption[]{The calculated Pb+Pb rapidity spectra \cite{LUCIFERI}
at $158$ GeV/c for $\pi^-$ and protons compared to measurements by NA49. 
The latter are for total $h^-$ and $h^+ - h^-$ respectively}
\label{fig:two}
\end{figure}

\begin{figure}
\vbox{\hbox to\hsize{\hfil
\epsfxsize=5.0truein\epsffile[24 85 577 736]{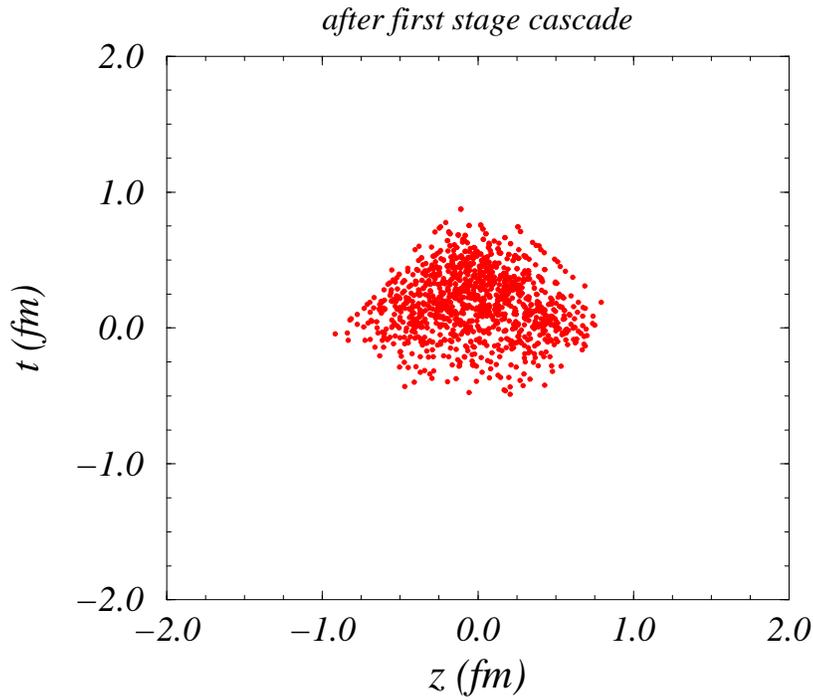}
\hfil}}
\caption[]{Positions of baryons in Pb+Pb after first phase: Fast cascade. The
time and longitudinal envelope indicates where the purely baryonic system
reaches after the two massive nuclei pass through each other at this SPS
energy $200$ GeV/c. Although this envelope represents the end of the fast 
cascade, some charmonium-baryon collisions may take place later, {\it i.e.\ }in the
comover phase.}
\label{fig:three}
\end{figure}

Nevertheless, calculations with a purely hadronic cascade \cite{LUCIFERI}
very well describe the energy loss and inclusive pion spectrum seen in
massive Pb+Pb collisions at SPS energies, see Fig(\ref{fig:two}).  These
apparently contradictory features in fact can be united into a resonance
based multi-scattering scheme.  Following the high energy cascade stage in
which collision histories are recorded and hard processes engage, the cascade
is reinitialised and a second hadronic cascade is carried out at greatly
reduced energy. Figure \ref{fig:three} shows the final positions of baryons
in the first phase, and the almost light-like paths for the particles
engaging in the initial cascade.

In the reinitialisation, the nucleon-nucleon interaction history, together
with the entire trajectory of each participant, is used to set up groups of
nucleons which have mutually interacted in the first stage. The
experimentally known averages, and the fluctuations inherent in $NN$
scattering, in energy loss, multiplicity, and character (flavour, {\it etc.})
of produced mesons are all used to produce additional generic mesons
associated with each nucleon group. Momentum, charge, baryon number and
flavour conservation are imposed on each group of baryons and associated
mesons.  These generic mesons are the principal cascaders in the second step,
along with the baryons.

\subsection{First Phase: High Energy Collisions}

As indicated, the procedure used here is relatively straightforward and in
outline resembles the eikonal or Glauber calculations made by previous
researchers \cite{Kharzeev,Gavin,Huefner}, but retains the random,
fluctuating, collisional nature of a cascade. This stage serves to establish
the space-time geometry of the interactions between the target and projectile
nucleons. Any actual hard processes which occur in this stage result in
real energy loss, but soft processes do not occur yet, they are delayed
until the reinitialisation. So, if a Drell-Yan pair is produced, its energy
is immediately subtracted, similarly for a $c\bar c$ state or an open charm
pair (we do not produce open charm at the moment, except by breakup of hidden
charm mesons). Clearly jets or mini-jets (partons) could be produced in the same way
and their hard evolution followed up to hadronisation, when the products
would simply be included during the reinitialisation before the second phase.

In practice, since Drell-Yan and J/$\psi$ production are very rare processes,
it is not necessary to actually globally conserve energy in making them, and
this would in any case lead to considerable difficulty since the basic
production must be considerably increased artificially to make calculations
possible in a finite time.  Therefore, energy is conserved in the first stage
in the sense that realistic rapidity and $p_t$ (and mass for Drell-Yan pairs)
distributions are employed for the appropriate elementary $NN$ collision
energy. 

Needless to say, the first stage interactions of charmonium preresonant states
produced in the first stage cannot be neglected. This is because the $c\bar c$
pairs are produced relatively quickly, and may undergo high energy collisions
with target or projectile nucleons still in the way. Therefore these
interactions are also counted for each charmonium, though actual
breakup or feeding to another channel is left until the reinitialisation.

The two phase approach has a highly beneficial effect on frame dependence of
the cascade, which has been touted by some authors as a possible limiting
factor on the utility of cascade models for the description of
ultra-relativistic ion collisions.  Preliminary calculations at RHIC energy,
$\sqrt{s}=200$, for a central Au+Au collision, in our context a worst case
scenario, yield at most a 10-15\% effect on $(dN/dy)_{\pi^-}$ between the
extreme cases of a cascade run in the global LAB frame and one in the global
CM frame.  Frame dependence at the SPS, to which energies the present work is
limited, is negligible within statistical errors. The reason for the great
reduction in frame dependence is simple and generic. The first, high energy
cascade leads, see Fig(\ref{fig:three}), to particles travelling uniquely on
light cone paths. Any residual frame dependence then enters only through the
second phase cascading, which involves greatly reduced energies and is not
very large. These results suggest that while frame dependence might be a
limiting factor on pure hadronic cascades at LHC, it is certainly not a major
difficulty at RHIC or SPS. But even partonic cascading is subject to similar
problems, through the use of albeit reduced but still finite cross-sections.

\subsection{Reinitialisation}

The fast cascade history is used to set up initial conditions for the second
low energy cascade. To begin, one needs positions and momenta for the baryons
and for the mesons {\it expected} to be produced from the initial
baryon-baryon collisions. As we have noted already, the baryons are formed
into groups which have collided with each other in the first stage, The
structure of the groups is virtually dictated by consideration of the p+A
system, where the projectile proton collides successively with those target
nucleons which are directly in its path.  

This simple grouping can be easily generalised to A+B collisions by using a
procedure which kinematically symmetrises each group with respect to target
and projectile.  One first selects a nucleon which had the maximal number of
collisions from all nucleons, both target and projectile, and adds to the
group all those nucleons with which it collided.  Since only high energy
collisions are involved in the first stage, these colliders are all going (in
an equal velocity frame) in a direction `opposite' to the originally chosen
particle. In p+A one is finished at this point. For A+B one proceeds to
choose from among the `opposite' colliders just added that nucleon which
itself had the maximal number of collisions. All the nucleons with which it
collided are finally added to the group.  Although geometry has carefully
been ignored in this construction, it clearly plays its accustomed role. 

There is no doubt some limited freedom involved in constructing the groups.
Quantitatively however, at least some of the possible alternative methods
make little difference to the final results. This comes about not least of
all because the final cascade phase really occurs at much lower energy, even
in much higher energy RHIC collisions. The model resembles an extension of
the wounded nucleon model\cite{woundednucleon}. We have essentially marked
each incoming nucleon with its multifaceted history, and one might refer to
this as a `painted' or programmed nucleon model.

The present work adds to the interaction choices the possibility of 
producing and destroying hidden charm $c \bar c$ states in each of the two
steps of the overall cascade. The production is accomplished almost totally
in the high energy phase, and uses elementary production cross-sections
normalized to pp measurements. Breakup is done using the collision history
for the charmonium collected in the first stage, through an interaction
matrix described in the section on coupled channels.  These processes are
constrained by pp and p+A measurements \cite{E772,NA50a,NA38}.

\begin{figure}
\vbox{\hbox to\hsize{\hfil
\epsfxsize=5.0truein\epsffile[24 85 577 736]{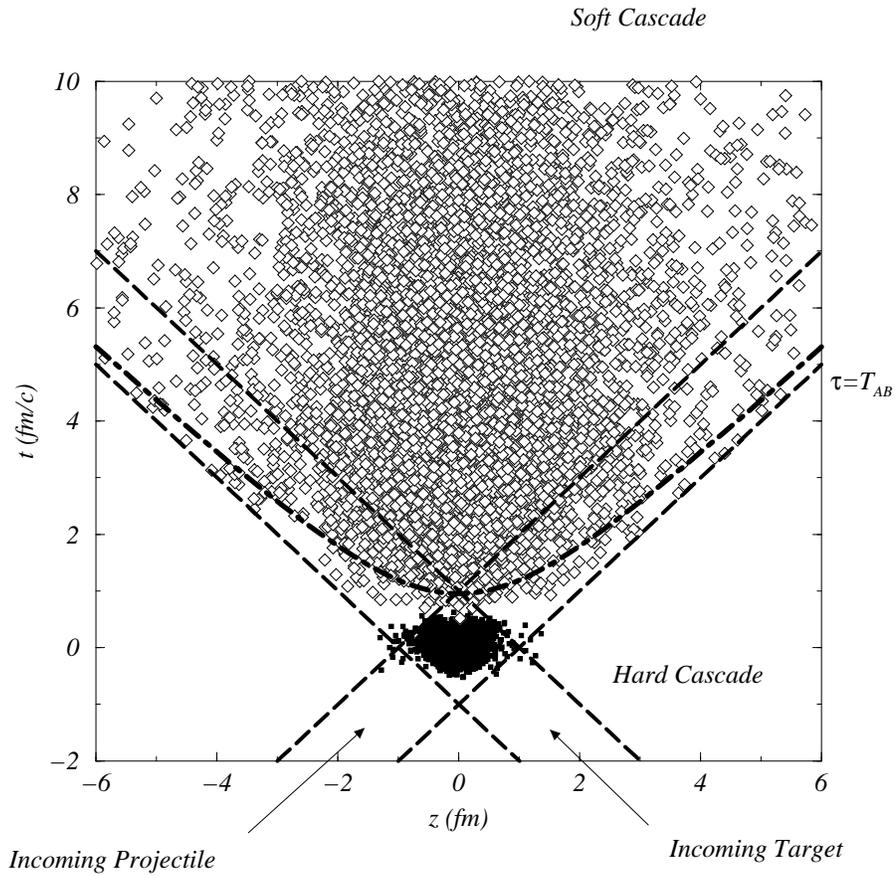}
\hfil}}
\caption[]{Time evolution of the ion-ion collision. The distribution in
space-time of collisions and decays in hard and soft cascades is shown
for the minimum bias Pb+Pb system. The initial paths of the incoming
nuclei are close to light-like and result in the dense initial blob
of hard collisions. The soft cascade occurs after the formation time
of mesons has passed, as indicated roughly by the constant proper time
surface.}
\label{fig:collisions}
\end{figure}

The final step in the reinitialisation places mesons and baryons in position
and time to restart the cascade. The four-momenta, in the global frame, of
all particles are known, as are the space-time coordinates of the initial
nucleons. It was thought best to distribute mesons produced within a group
randomly along the paths of the baryons in that group.  This choice would
seem reasonably consistent with locality principles. In any case, sins
committed in this way are remediated by the formation time that must be
attributed to each meson before it can begin to interact in the last phase.
See Fig(\ref{fig:three}) for the final distribution of baryons following the
hard cascade, and Fig(\ref{fig:collisions}) for a spacetime picture of the
collisions in the entire cascade, both soft and hard phases.
\subsection{Soft Phase}

The second phase then involves generic resonances, baryonic and mesonic, like
the $\Delta$, $N^*$, $\rho$ and $\pi$ in quantum numbers.  As stated, these
have masses between $M_N =0.939$ GeV and about $2$ GeV for baryons and from
$0.3$ GeV to $1.1$ GeV for non-strange mesons, and appropriately higher for
strange mesons. These produced resonances are not allowed to interact until a
sufficient formation time has transpired.  This time, $\tau_f$, is a {\it
real} adjustable parameter in the model, to be fixed perhaps from p+A, or as
in earlier work from light nucleus collisions (S+S) \cite{LUCIFERI}.  We
imagine that these are broad $s$-wave excitations of the underlying
representatives, not series of special and very narrow states such as are
tabulated in the particle data book. These narrow states cannot be excited
very much in actual ion-ion collisions, and do not carry much of the $pp$
cross-section. Finally, these generic resonances decay by sequential pion emission
into lower mass excitations, losing mass with each decay until physical
$\pi$'s, $\rho$'s, and $K$'s, or stable baryons are the only open channels.

The decay time for the generic excitations is a second real parameter,
perhaps to be taken inversely proportional to the excitation mass, but here
for simplicity fixed at $\tau_d \sim 1/125$ MeV$^{-1}$. It is often assumed
that one's lack of knowledge of resonance-resonance scattering opens up to
cascade models a deep well of adjustable parameters. This is not the case
here. We employ, as did Gottfried \cite{gottfried}, a `universality'
principle for soft interactions. Surely, for soft baryon-baryon interactions
all fine details, excepting perhaps size and mass thresholds, are irrelevant
in determining what must be, basically, interactions driven by many gluon
exchange. We ignore size differences for the moment, with the exception
later, of such very small objects as charmonium mesons. This limits the
number of free parameters in the model to a minimum, in fact only a few, so
far the two times $\tau_f$ and $\tau_d$. The important $pp$, $p\bar p$ and
$\pi p$ data, determined from experimental measurements over a wide range
energies at which cascading takes place, are the primary inputs, and serve to
fix all of the baryon-meson and baryon-baryon interactions.  Meson-meson
interactions are fixed by appealing to constituent quark model counting.

\section{Model for Hadron-Hadron Interaction}

The objective of the cascade approach to ion-ion collision is to proceed from
a knowledge of elementary hadron-hadron collision to a prediction of the far
more complex many body event.  Many approaches have been put
forward \cite{wang,RQMD,URQMD} including strings \cite{frithjof,werner}, but we
prefer to retain a particle nature for the cascade.

\begin{figure}
\vbox{\hbox to\hsize{\hfil
\epsfxsize=5.0truein\epsffile[31 81 579 702]{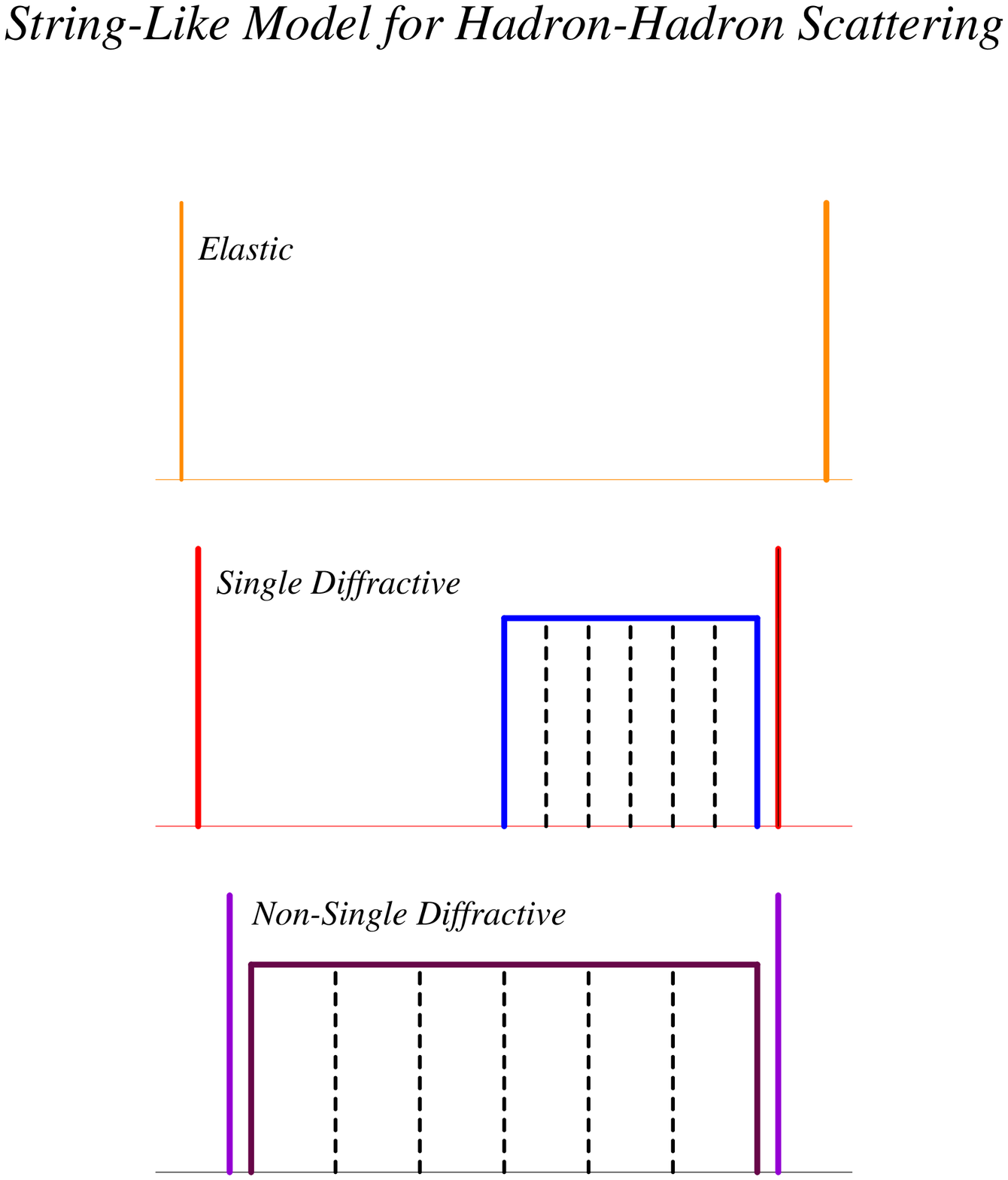}
\hfil}}
\caption[]{Shown are graphic representations of the elements of the model
for the elementary hadron-hadron collision: elastic, single
diffractive (SD) and non-single diffractive (NSD). The meson groups
introduced in both SD (with a rapidity gap) and NSD have a string -like
character but divided into our generic resonances. It is customary to
associate SD with a three pomeron coupling and NSD with one, two or more
pomeron exchange.} 
\label{fig:four}
\end{figure}

The required input is a model for the elementary hadron-hadron system,
beginning with nucleon-nucleon but easily extended to meson-nucleon and
ultimately applied to any two body hadron-hadron collision. The basic
processes are elastic scattering and inelastic production of mesons.  The
latter we divide into the well known categories \cite{Goullianos}:
diffractive scattering, referred to as single diffractive (SD), and
non-single diffractive (NSD) \cite{UA5}. A graphic description of these
processes is given in Fig(\ref{fig:four}). The SD process, leading to a
rapidity gap between one of the leading hadrons and the produced mesons
cluster, is associated with the triple Pomeron coupling \cite{pomeron}, while
the NSD production is attributed to single (and double) Pomeron exchange and
presumably results in the observed meson plateau.  These diagrams are the
basis for our hadron-hadron model but must be supplemented by an intermediate
picture which allows us to apply them, not only to hadron-hadron interactions
in free space but also inside a nuclear environment. 

The generic mesons depicted in Fig(\ref{fig:four}) and generic baryons,
having rather light masses selected in the ranges suggested above, constitute
the basic elements for rescattering in the second-phase cascade. Our
principal phenomenological sources are tied to the rapidity \cite{Goullianos}
and multiplicity \cite{UA5,Blobel,hhisr,FNAL} information obtainable from the
elementary collisions. Care must be taken to describe the observed
multiplicity distributions. We have chosen to impose KNO
scaling \cite{LUCIFERI,KNO} in our parameterisation of these distributions,
despite the relatively small deviations seen for example in the CERN UA5
experiments \cite{UA5}. We use the distributions obtained in
reference \cite{LUCIFERI}, covering a wide range of energies from a few GeV to
100's of GeV for pp and finally to $200$ GeV/c and $1.8$ TeV/c for $p\bar p$
\cite{UA5,FNAL}.

\section{Coupled Channel Model for Charmonium}

\begin{figure}
\vbox{\hbox to\hsize{\hfil
\epsfxsize=5.0truein\epsffile[24 85 577 736]{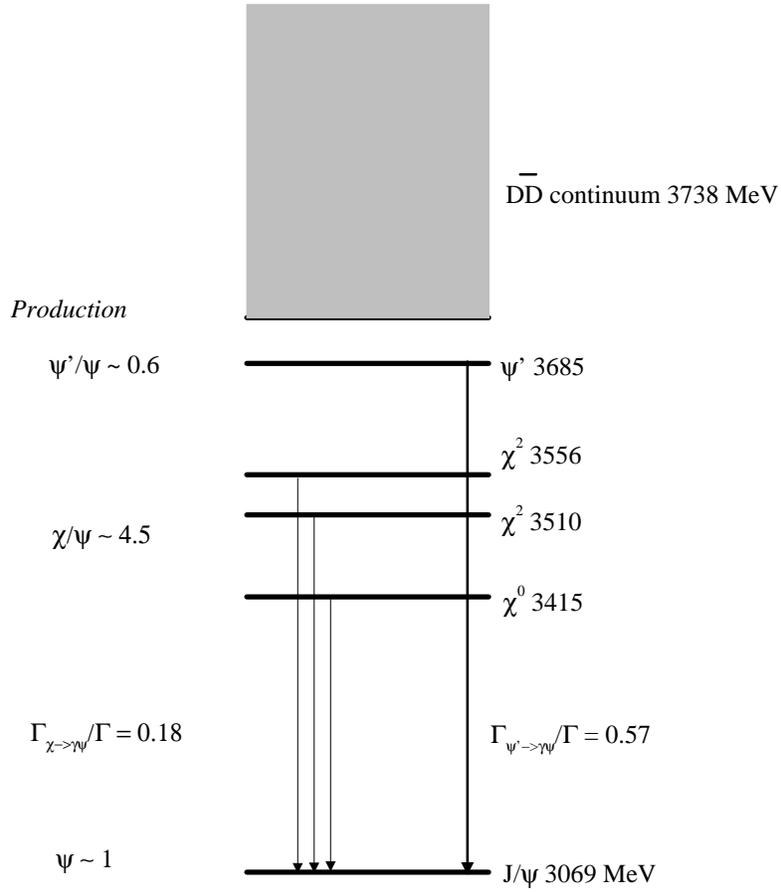}
\hfil}}
\caption[]{Charmonium spectroscopy including higher mass states which are
significantly produced in $pp$ and which feed strongly to the J$\psi$.
Electromagnetic and hadronic decays of $\chi^i$ (a weighted average) and
$\psi'$ are both included in the indicated branching ratios. The production
ratios are suggested by direct measurement.} 
\label{fig:levels}
\end{figure}

The treatment of the hidden charm {$c\bar c$} mesons within a `purely'
hadronic code presents some problems, perhaps not fully solvable within the
effective hadronic treatment of such states. We do not deviate much in spirit from
the work of previous researchers \cite{Huefner,Kharzeev,Gavin}, but the devil
lies sufficiently in the details to produce some quantitative effects. The
production of charmonium mesons is almost completely limited to that coming
from nucleon-nucleon collisions at the highest energies, {\it i.e.\ }in the
initial high energy cascade, not by fiat but by the greatly reduced collision
energies in the second phase. Destruction of the charm meson precursors, in
contrast, can occur in the first baryonic phase and also later in collisions
with generic mesons and baryons in the second, low energy phase, {\it i.e.\
}on comovers. It is in the destruction of the charmonium states we differ most,
ascribing a more direct role to the presence of the higher mass $\chi$ and
$\psi'$ mesons, for which in fact breakup is far easier. We include in
Fig(\ref{fig:levels}) a level diagram showing the relevant charmonium states
to make the picture as clear as possible.

In the actual calculations of the above cited references both production and
breakup are treated as instantaneous. There is no J/$\psi$ formation time in
the high energy phase or its near cousin, the Glauber or eikonal
modeling. Kharzeev {\it et al.\ }\cite{Kharzeev} in fact justify such a
choice by referring to microscopic production of charmed quark pairs, the
subsequent formation of a preresonant-resonant state from which all charmonium mesons
emanate, and the breakup in relatively hard scattering by gluons radiating
from nearby nucleons. All of this can be incorporated in the initial cascade
by ascribing to $NN$ collisions a production cross-section normalized to the
measured elementary $pp$ and $pn$ data and similarly a baryonic breakup
deduced from say p+A measurements. Nevertheless, the eventual result is an
effective hadronic modeling for the charmonium states, ascribing to a produced
$c \bar c$ pair a hadronic state which can be destroyed.  

We imagine that the primordial $c\bar c$ pairs are originally produced
essentially in plane wave states. Clearly, both singlet and octet color
states are involved. This view would seem to be a reasonable one given the
predominance of open charm production over hidden charm production in free
space $NN$ collisions. We further suppose that in elementary $NN$ collisions
the {$c\bar c$} pair eventually coalesces, with a state dependent
probability, into a J/$\psi$, $\psi'$ or $\chi$. The time which elapses will
be determined by the size of the bound state and the probability that a
transition occurs. The probability of formation will depend critically on the
relative momentum of the coalescing pair as well as on their spatial
separation. In any picture of charmonium generation there must needs be some
formation time $\tau_b$ for the bound state, which may be, in general, longer
than the total duration of the fast, baryonic cascade $T_{AB}$.  Then, in an
ion-ion collision, additional transitions may be induced: into the continuum,
{\it i.e.\ }breakup, or perhaps also between bound states.

Therefore, whether one sees the early evolution of the eventual charmonium as
a preresonant state or as plane waves may be immaterial. Given the small size of
the J/$\psi$ as opposed to the much larger $\psi'$ and $\chi$, the separation
of the $c$ and $\bar c$ in the plane wave picture could equally well serve as
a distinguishing feature. What can differentiate our calculation from the
earlier models is the possibility of transitions between charmonium states
dynamically in a nuclear environment. Certainly, the $\chi^1$, $\chi^2$ and
$\chi^3$ states are produced considerably more copiously in basic $pp$
collisions, with perhaps as high a ratio as $\chi/\psi = 4-5$ \cite{isr-chi},
and they decay appreciably into J/$\psi$, with branching ratios in the range
$\Gamma_b/\Gamma \sim 12-25$\% \cite{bluebook}.  The $\psi'$ also feed some
$57\%$ into J/$\psi$. It follows that one cannot ignore their presence.

This point becomes even more significant when one considers what the breakup
probabilities for the higher mass charmonium states are likely to be, either
in the fast or slow cascades. These heavier objects are considerably larger
spatially and might well have total cross-sections on baryons or mesons
proportional to the square of their colour dipole radius \cite{Huefner2}. In
any case, in the precursor states of the initial rapid cascade the spatially
larger charmonia will have more time for collisions before forming, and
should then possess larger effective breakup strengths.  It is interesting
that the J/$\psi$ production cross-sections on baryons and on mesons differ
little, perhaps also a reflection on annihilation. We interpret the observed
suppression of J/$\psi$ as a measurement of the $\psi'$ and/or $\chi$
strength as well. One of our conclusions will be that a considerable portion
of the anomalous suppression seen in Pb+Pb, even for quite large impact
parameters, is a result of breakup in the higher charmonium states and an
extinction of their free space feeding down.

What extra parameters has our model introduced relative to other treatments?
We introduce  a reaction matrix for charmonium states $R_{ij}$ which permits
transitions between the states as well as diagonal, breakup,
elements. Unitarity constrains this matrix. The diagonal elements should be
present for other practitioners also but are not, in general, since only one
preresonant charmonium state is usually considered.  The only significant
off-diagonal component for our purposes is that for the $\psi$ to $\psi'$
transition, shown below to be small, but potentially influential. The quantitative
reaction matrix $R_{ij}$ is specified in the ensuing sections treating the
explicit calculations. Indeed, in our ``standard'' calculation, encompassing
all major results presented below, this matrix is limited to diagonal in
form, thus reducing the number of adjustable parameters.  The breakup matrix
elements for the second phase cascade, {\it i.e.\ }for breakup on comoving
mesons, are scaled from the first phase by counting constituent quarks.

The formation time for secondary mesons, $\tau_f$, controls the initiation of
the second cascade and thus the onset of comover destruction of charmonium
and the density of comovers. A reasonable choice for this parameter is
$\tau_f\sim 0.5-1$ fm/c, and this is in fact consistent with the production
of $\pi$ mesons in ion-ion collisions at the SPS \cite{LUCIFERII,NA49,NA35}.
The high density of comovers which obtains at these times implies they play a
considerable role. The effective formation time is actually somewhat longer,
since it is increased by the duration of the fast cascade,
{\it i.e.\ } $\tau_{eff}=\tau_f+T_{AB}/2$.

\begin{figure}
\vbox{\hbox to\hsize{\hfil
\epsfxsize=5.0truein\epsffile[24 85 577 736]{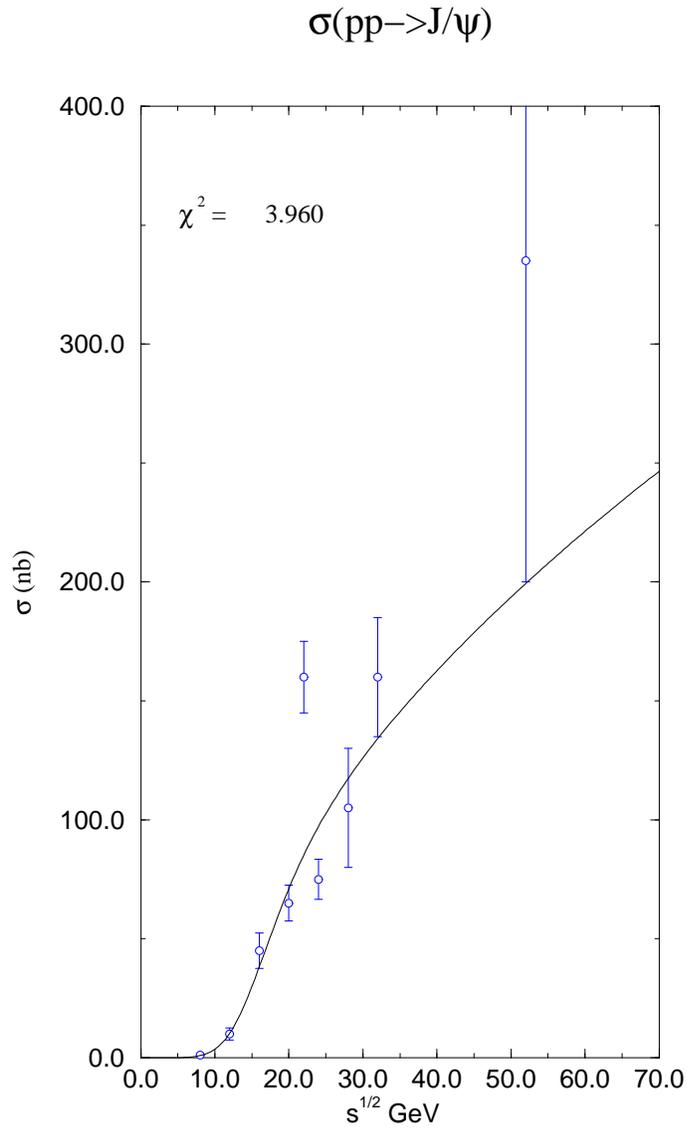}
\hfil}}
\caption[]{Production of J/$\psi$ from pp as a function of energy. The 
$\pi p$  cross-section is also known, and in fact is very similar to that for
$pp$, but rarely plays a role with production generally significant only 
at the highest energies.} 
\label{fig:jpsi-xs}
\end{figure}

The energy dependence of the elementary J/$\psi$ production cross-sections is
shown in Fig(\ref{fig:jpsi-xs}).  The sharp dependence of $\sigma_{J/\psi}$
on energy near the SPS values $\sqrt{s}= 17-20$ GeV implies that virtually
all production occurs in the high energy phase. Drell-Yan exhibits a similar
behaviour.
 
\section{Drell-Yan}

The high energy phase, designed to record the initial interactions of the
nucleons in the two colliding nuclei also provides the basis for our estimate
of massive dilepton production. {\it i.e.\ }Drell-Yan, an important side of
the quandary we faced at the start. We limit ourselves to the canonical FNAL
\cite{E772} p+A measurement at 800 GeV/c, but in fact the method of
calculation guarantees agreement with the lower energy p+A and A+B collected
by NA50\cite{NA50a}. The calculation is straightforward and was a first
example, for us, of a hard process added to the initial high energy
cascade. Drell-Yan is generally considered to be calculable perturbatively
for dilepton pairs with masses in excess of $M_{\mu\mu}=4$ GeV. Production in
the short time defined by such masses proceeds without energy loss and leads
to the $A$-dependence shown in Fig(\ref{fig:one}). The E772 results
\cite{E772} imply very close to linear $A$ dependence. To perform the
Drell-Yan microscopically we have introduced parton structure functions
\cite{LUCIFERI}, and of course the requisite parton variables, into the
code. But the curves in Fig(\ref{fig:one}) could have been obtained purely
geometrically from the elementary production rates and the high energy phase
only; very little production comes from the second, low energy phase. Any
cascade which does not correctly describe this feature of Drell-Yan is in
danger of producing spurious charmonium suppression by means of premature
energy loss.  Drell-Yan and J/$\psi$ production both can occur in second,
third and higher order collisions of initial nucleons as well as in the first
collision, and if energy is lost by the nucleons immediately following each
collision, the result will be $A$-dependent `suppression', since the
production of all charmonium states drops sharply with decreasing
energy. 

Later we introduce for purposes of comparison a survival probability for
J/$\psi$, which has as its denominator, aside from a nucleon-nucleon
normalisation, the Drell-Yan yield corresponding to the charmonium yield in
the numerator. Thus, apart from overall normalisation the ratio
$\sigma(\psi)/\sigma(DY)$ and the J/$\psi$ survival probability $P_s =
N_\psi(survived)/N_\psi(produced)$ are one and the same thing.  This is
because {\it elementary} Drell-Yan production and J/$\psi$ production are
treated in exactly the same way in our cascade, and  the number of high mass
lepton pairs tracks the {\it initial} number of charmonia.

\section{Charmonium Suppression in Nuclear Collisions}

\subsection{Minimum Bias: p+A and comparison to Glauber}

We begin with the suppression in p+A for which comovers play little role, the
product pA of atomic numbers is small in comparison to say Pb+Pb. Even here,
however, the first stage high energy cascade does not suffice for an accurate
description, some of the suppression on baryons occurs only in the second
stage, as slow J/$\psi$'s emerging from the interaction region are caught by
nucleons, or interact at low energy in the target. The nucleon-nucleus data
provides a necessary constraint on the basic parameters to be used in
baryon-baryon production and breakup. This simplified system also provides a
fruitful ground for comparison between the eikonal approach and LUCIFER.

\begin{figure}
\vbox{\hbox to\hsize{\hfil
\epsfxsize=5.0truein\epsffile[24 85 577 736]{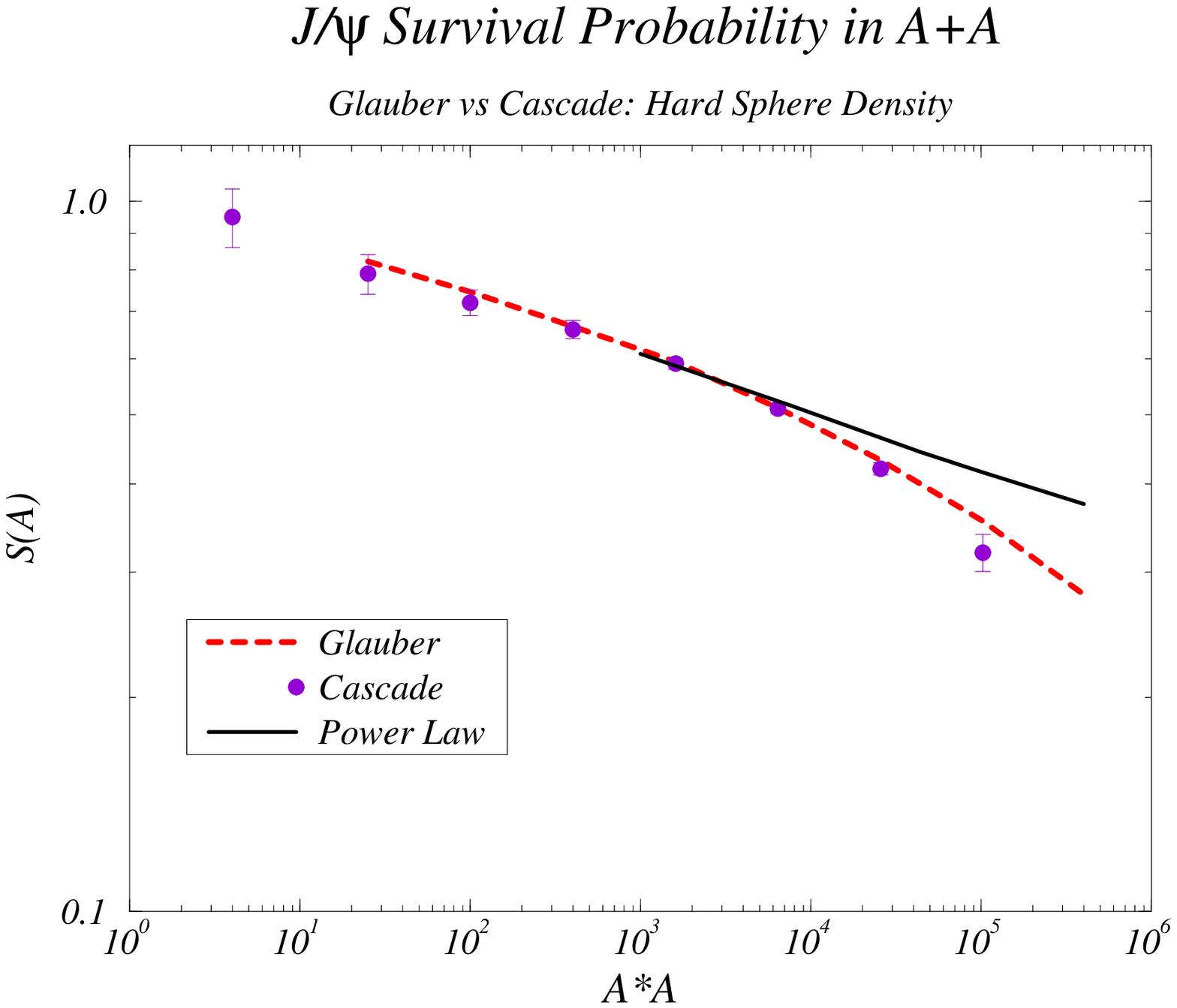}
\hfil}}
\caption[]{Comparison for A+A between Glauber and cascade, the latter in a
purely J/$\psi$ mode and both calculations employ $\sigma_br=7.3$ mb. The
deviation from a power law is apparent for large A$\times$A. A hard sphere form is
used for the nuclear density.} 
\label{fig:Glauber}
\end{figure}

To facilitate a comparison  with the cascade we have made our own 
calculations with the Glauber formalism, relying on the formula:

\begin{eqnarray}
\frac{dS_{Gl}}{d^2 b} & = & \frac{1}{AB\,\sigma(NN\rightarrow\psi)}
\left\lbrack
\frac{d\sigma(AB)}{d^2b}
\right\rbrack \\
& = & \int\,d^2s\,dz\,dz'\,\rho_A(s,z)\,\rho_B(b-s,z')
\,I_A(s,z)\,I_B(b-s,z'),
\end{eqnarray}
\begin{equation}
I_A(s,z) = 
\left\lbrack 
-(A-1) \int_z^\infty\,dz_A\,\rho_A(s,z_A)\,\sigma_{abs})
\right\rbrack,
\end{equation}

\noindent for the differential survival probability of J/$\psi$ produced in
p+A collisions, with all integrals, including the one over $b$, carried out
numerically to obtain the total survival probability.  Here $\sigma_{abs}$ is
the J/$\psi$ breakup cross-section and is to be determined from p+A data. We
follow Reference\cite{Kharzeev} in this development but employ a simpler,
hard sphere, version of the nuclear density $\rho_A(b,z)$ for the purpose of
comparing Glauber and LUCIFER II. It was instructive to extend this
comparison to A+A collisions to demonstrate that even Glauber does not
reproduce the canonical power law, implied in the experimental descriptions
\cite{NA50a,NA38,NA49} which always are compared with a straight line fit on
a log-log plot, supposedly arising from purely baryonic breakup.  These
results are displayed in Fig(\ref{fig:Glauber}) for the J/$\psi$ without
coupled channels to make an easier comparison between cascade and Glauber.
The J/$\psi$ absorption cross-section is taken so as to reproduce the p+A
observations at $800$ GeV/c \cite{E772}, $\sigma_{abs}\sim 7.0$ mb with the
hard sphere configuration. But this value is equally successful for the lower
SPS energies.

\begin{figure}
\vbox{\hbox to\hsize{\hfil
\epsfxsize=5.0truein\epsffile[24 85 577 736]{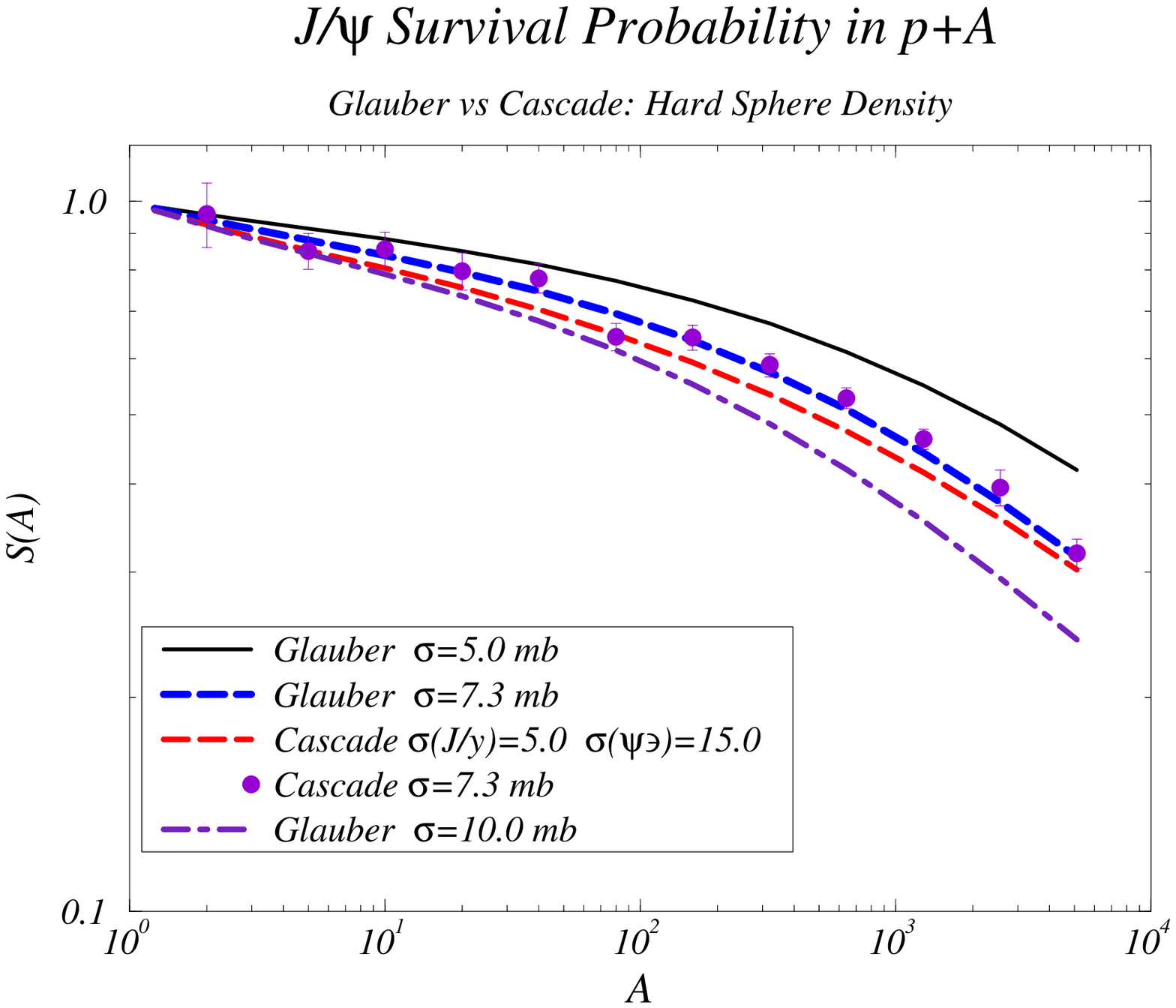}
\hfil}}
\caption[]{Again comparison of survival probabilities for Glauber and
cascade, the latter appearing both in pure J/$\psi$ and coupled channel
mode. A variety of absorptive strengths are illustrated. The use of a hard
sphere density distinguishes these results from later cascade simulations; in
particular leading to small changes in the breakup cross-sections.}
\label{fig:Glauber2}
\end{figure}

A second comparison, again for minimum bias production of $J/\psi$, appears
for p+A in Fig(\ref{fig:Glauber2}). In this we use the coupled channel
modeling, whose details we elaborate further now.  The relative production of
the different charmonium states is taken so as to reproduce the pp data from
the ISR \cite{isr-chi} for the $\chi$ to J/$\psi$ ratio, {\it i.e.\
}$\chi/\psi\sim 4.5$, and for appropriate $\psi'$ production
\cite{pp-psip}. A general ballpark for the measured $\chi$ to J/$\psi$
production ratio in pp is a factor of $4-5$ \cite{isr-chi,pp-psip} and our
final results are rather insensitive to a choice in this range, since a
decrease could easily be compensated by a small transition matrix element
between J/$\psi$ and $\chi$.
 
The $\psi'$ to J/$\psi$ production in pp is taken near $0.33$ so as to
reproduce the final J/$\psi$ contribution from the eventual decay of the
$\psi'$ seen in a variety of experiments at a range of energies
\cite{NA50a,pp-psip}, {\it i.e.\ }implying an eventual $\psi'$ to J/$\psi$ ratio
$\sim 0.15$. The free decay of the $\psi'$ into J/$\psi$ is of course given
the standard \cite{bluebook} value $0.57$.

The $\chi^1$ and $\chi^2$ (and all other charmonium states) are assigned their
correct masses, to properly include threshold effects in breakup. But the
branching into J/$\psi$ is taken the same $0.18$, as a weighted average over
the measured electromagnetic values together with a very small hadronic
component, $\le 0.5\%$ \cite{bluebook}. Again, small variations $\sim 1-2\%$
in this branching ratio have little effect on suppression in any of the charmonium
states and can be compensated for by commensurate changes in in, for example,
overall cross-section normalisation.

A first inference to be drawn from Fig(\ref{fig:Glauber}) for A+A and
Fig(\ref{fig:Glauber2}) for p+A is that the first high energy cascade
produces pure J/$\psi$ dynamics very much like that in Glauber.  For the
above choices a $\sigma_{abs}\sim 7.0$ mb leads to very nearly the same yield
with the cascade as with Glauber.  A second lesson, key to our development,
is that the coupled channel model reproduces the Glauber result for J/$\psi$,
using a smaller direct breakup cross-section, here taken as
$\sigma_{abs}(J/\psi)= 5-6.0$ mb, and including indirect destruction via the
considerably larger $\sigma_{abs}(\chi)=3\,\sigma_{abs}(J/\psi)$ for $\chi$ and
perhaps higher for $\psi'$. The increased spatial sizes of the higher states
strongly support the use of larger absorption cross-sections. A final
observation for collisions extending to $A\times B$ values encompassing S+U
and further to Pb+Pb, is that pure Glauber theory and the first stage nucleon
cascade, both produce lines curving appreciably downwards on log-log plots,
deviating from any power law. One gathers there is a little bit of
`anomalous' suppression even in a bare bones, no comover, theory.

One should keep in mind that the true charmonium states in p+A collisions are
produced mostly outside the nucleus and we and other workers are, for this
first cascade, considering charmonium progenitors, perhaps preresonances or
perhaps just comoving $c\bar c$ pairs at a certain separation. The effect is
however the same as using effective charmonium states, instantaneously
produced, as we and previous researchers \cite{Huefner,Kharzeev,Gavin} are
doing.  This situation is altered when one turns to the second stage cascade which
begins later, when all mesons, charmonium and others, may have had time to
precipitate.

\subsection{Suppression in A+B Collisions}

We have, in the last section, already presented results of simulations for the
high energy cascade alone, and demonstrated the close similarity of this
first stage to an eikonal approximation. The p+A results,
compared to both Glauber and experiment, were displayed only for minimum
bias collisions. To complete the picture one must allow the soft cascade to go forward,
especially for ion-ion collisions where the production of mesons becomes very
significant. A key parameter in second cascade is the delay, $\tau_f$ or more
properly $\tau_{eff}$, afforded by the time scale involved in the `soft'
formation of mesons.  We reemphasize that we use a more or less standard
value $\tau_f \sim 0.5-1.0$ fm/c, but are constrained by the production of
mesons in the most massive system considered here, Pb+Pb. This production
was considered extensively in reference \cite{LUCIFERII}. Perhaps a $10-20\%$
uncertainty might attach to $\tau_f$.

There are two sets of data to be considered: first, minimum bias J/$\psi$
cross-sections as a function of the product $A\times B$ of nuclear atomic
numbers, and second the ratio of J/$\psi$ yield to Drell-Yan yield as a
function of centrality, or more specifically transverse energy $E_t$. 

Our results for minimum bias are displayed for the combined effect of both
cascade phases in Fig(\ref{fig:jpsi-minbias}). The anomalous suppression in
Pb+Pb is well reproduced by the totality of our two step, but otherwise
conventional, hadronic dynamics.  Part of the additional suppression in Pb+Pb
relative to S+U already arises from the high energy cascade, coming from the
increased $\chi$ and $\psi'$ breakup in the more massive nuclear
collision. But a considerable differential suppression arises from comovers,
some $40\%$ of the difference between S+U and Pb+Pb, the rest arising from
the hard cascade. Thus our conventional, hadronic explanation of the
interesting NA50 measurements is multifaceted, being rooted in both of the
two cascade stages, in the coupling of charmonium channels, and in the
presence of comovers. Part of the anomaly however, is perhaps illusory in
view of the `curving down' seen for large $A\times B$ in
Fig(\ref{fig:Glauber}) and Fig(\ref{fig:Glauber2}).

\begin{figure}
\vbox{\hbox to\hsize{\hfil
\epsfxsize=6.0truein\epsffile[24 85 577 736]{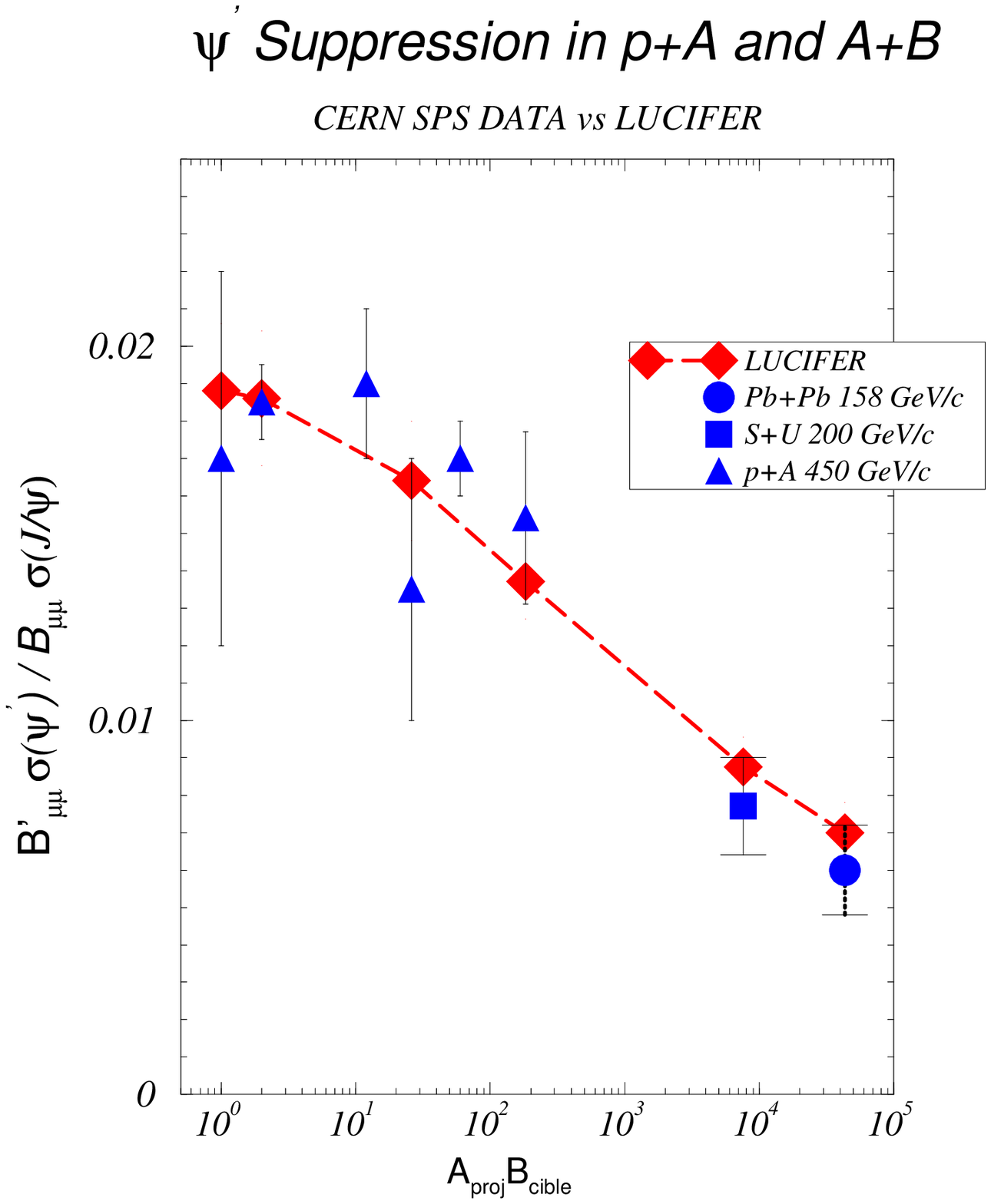}
\hfil}}
\caption[]{Comparison of experiment vs simulation for $\psi'$. 
The Pb+Pb data from NA50 was rescaled to 200 GeV/c by the collaboration. 
The S+U data is taken from NA38. The cascade calculations, again normalised
to nucleon-nucleon, reproduce the observed behaviour for p+A and the sharp
drop in the $\psi'$ to J/$\psi$ branching ratios for the massive nuclear collisions.}
\label{fig:psip-minbias}
\end{figure}

The calculated minimum bias $\psi'$ suppression is compared to data in
Fig(\ref{fig:psip-minbias}). The strong drop occasioned by the large increase
from p+W to S+U or Pb+Pb is clearly present in the theory. As
is evident in this figure  the $\psi'$ breakup strength inferred from p+A proves
sufficient for both S+U and Pb+Pb.

\begin{figure}
\vbox{\hbox to\hsize{\hfil
\epsfxsize=6.0truein\epsffile[24 85 577 736]{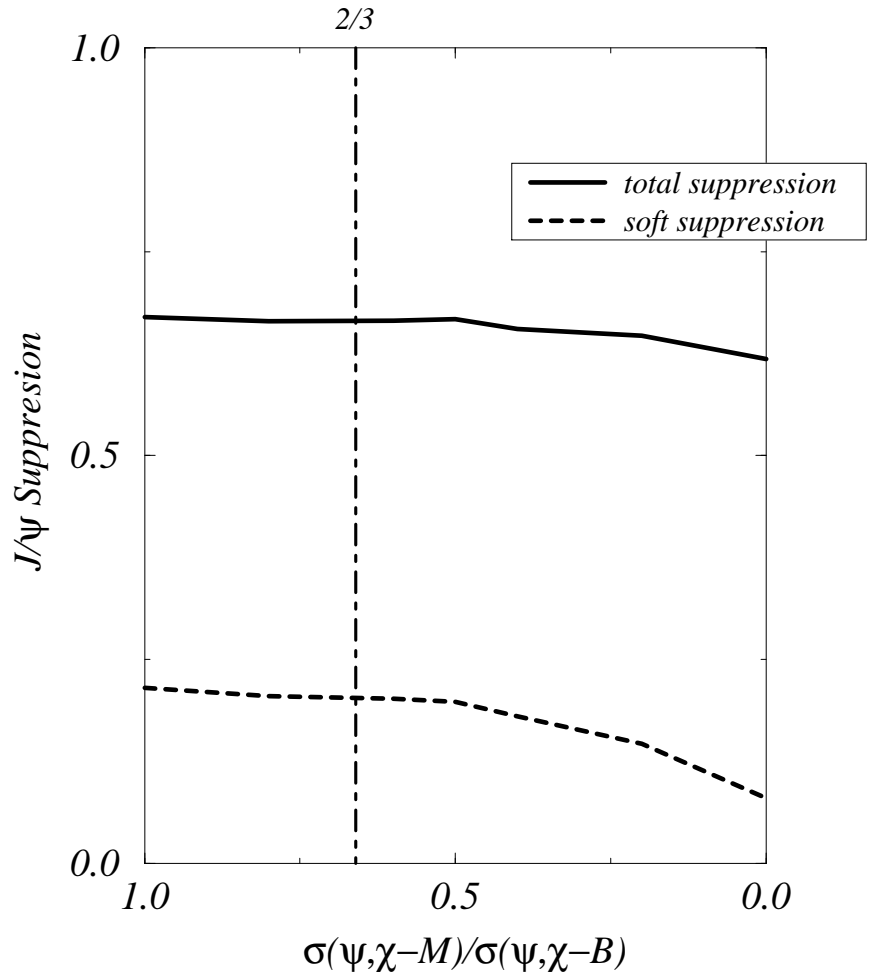}
\hfil}}
\caption[]{Variation of J/$\psi$ suppression with the charmonium-meson
cross-sections. We use 2/3 as the ratio to charmonium-baryon for the
calculations in the paper.}
\label{fig:comovers}
\end{figure}

The breakup cross-sections in these simulations are $6.6$, $20.0$ and $25.2$
mb for the $\psi$, $\chi^{i}$ and $\psi'$ respectively. These represent
absorption in charmonium-baryon collisions, and are reduced by the
constituent quark factor $2/3$ in $\psi$-meson. Variation of these
meson-meson cross-sections upwards to full equality with charmonium-baryon
leads to $\sim 0.5\%$ change in the overall J/$\psi$ suppression for
Pb+Pb (see Fig(\ref{fig:comovers})). This small change is muted by
two circumstances, even in the soft cascade the suppression from
charmonium-baryon collisions is still $1/3$, and also the presence of a
saturation effect in the degree of J/$\psi$ suppression. Hence, one obtains a
robustness in the predictions.

This perhaps surprising non-linearity for the J/$\psi$ interaction arises
because in our model this state is intrinsically tied up with the higher
states.  Introducing off diagonal elements $R_{ij}$ would produce a family of
solutions. We have left well enough alone; the present few modeling parameters,
now mostly determined independently of S+U or Pb+Pb data, surely having
produced already an adequate description of observation.

The survival probabilities for the J/$\psi$ in S+U are $0.50$ and $0.87$
in the hard and soft cascades respectively. The same figures for Pb+Pb
are $0.42$ and $0.775$. These results are both obtained using the model
parameters described above and an effective formation time $\tau_{eff}\sim
0.6+0.9/2.0$ fm/c, that is close to $1$ fm/c. Comovers play
a significant role, increasingly more so in Pb+Pb, but one must restate that
baryons as well as mesons represent comovers in the soft cascade. The ratio
of breakup on mesons vs that on baryons is $\sim 2:1$ for Pb+Pb and less for
S+U.

\subsection{Centrality: Dependence on Transverse Energy}

\begin{figure}
\vbox{\hbox to\hsize{\hfil
\epsfxsize=6.0truein\epsffile[24 85 577 736]{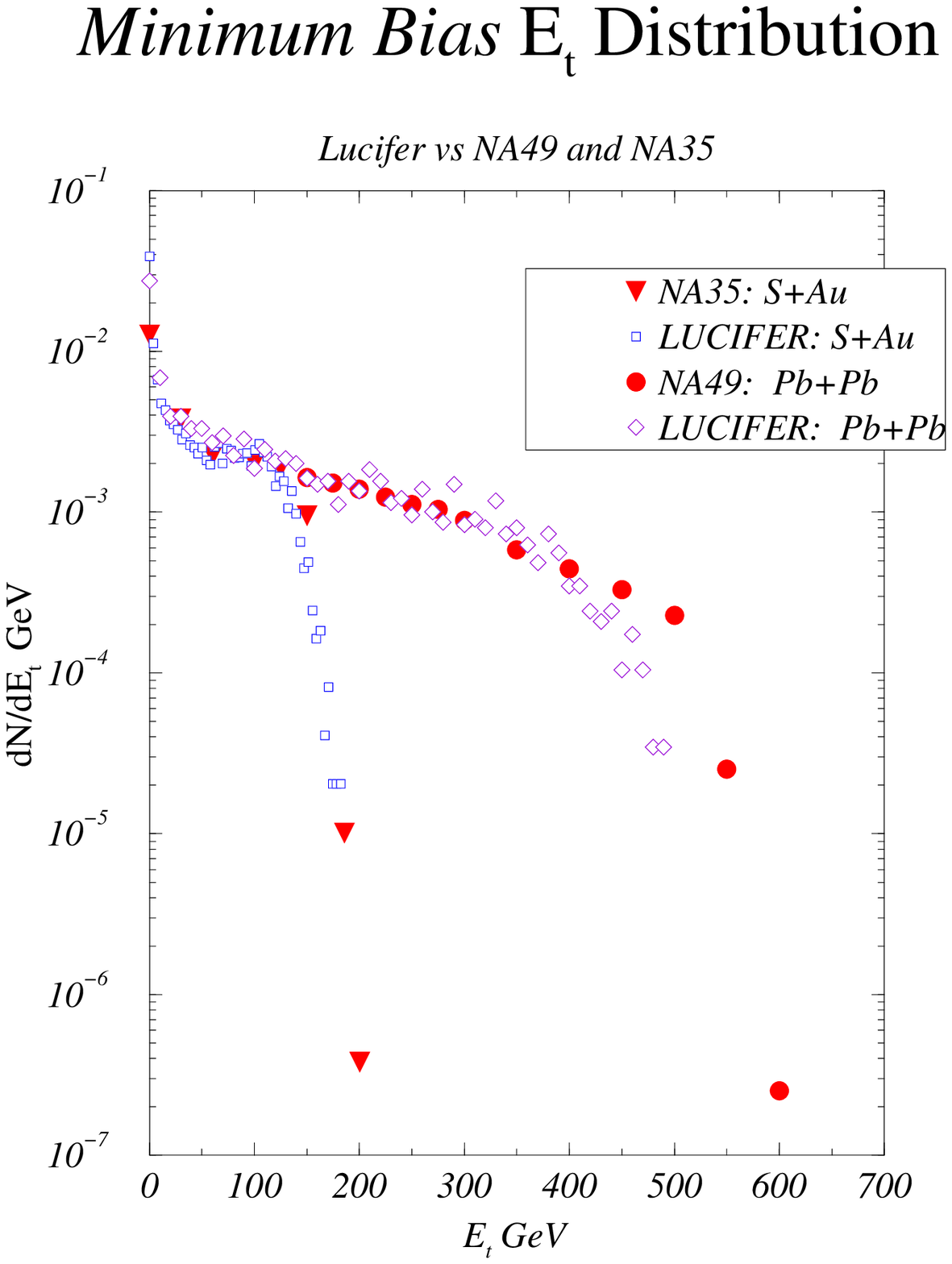}
\hfil}}
\caption[]{Transverse energy distributions from LUCIFER compared to 
experiment.}
\label{fig:ET-NA49}
\end{figure}

Perhaps the most striking features of the NA50 \cite{NA50b} measurements are
contained in their plot of J/$\psi$ suppression vs $E_t$. Unlike the existing
Glauber calculations of transverse energy the cascade provides a built in
$E_t$ scale, which does not necessarily agree exactly with the experimental
determination. There is no theoretical parameter to adjust the simulated
spectrum and the comparison of theory and experiment is a further test of the
cascade.  NA50 plots for J/$\psi$ and Drell-Yan show $E_t$ from neutral
mesons ($\times 3$ presumably in order to include all charge states) within
the pseudo-rapidity range $\eta=1.1$--$2.3$. To establish a calibration from
LUCIFER II we first refer to their earlier Pb+Pb results \cite{NA49} using a
more central rapidity range $\eta=2.1$--$3.4$, and including both
electromagnetic (neutral) and hadronic calorimeters to estimate $E_t$. This
comparison is shown in Fig(\ref{fig:ET-NA49}) \cite{NA49}, and indicates that
LUCIFER II, with standard parameters \cite{LUCIFERII}, provides a reasonable
representation of the measurements.  The small discrepancy between cascade
and experimental endpoints, some $10\%$, should be kept in mind when
examining the NA50 charmonium data, but is unlikely to have a major effect on
the results.

Charmonium breakup in the second phase cascade occurs both on baryons and on
mesons, generic and stable. Since meson numbers can be large, in particular
for Pb+Pb, their contribution to suppression may be appreciable. For
collisions of J/$\psi$ with generic mesons of mass greater than the
charmonium binding, {\it i.e.\ }the energy interval $\delta E\sim 600$ MeV from the
charmonium mass to the $D \bar D$ continuum, we employ a properly exothermic
cross-section. We use $\sigma_{abs}\propto k_f/k_i$ for the enhancement of the
reaction cross-section. The enhancement due to the inverse $k_i$ factor is
not large numerically however, yielding a couple of percent at most in the excess
suppression attributable to comovers.

Figures (\ref{fig:SUcentral}) and (\ref{fig:PbPbcentral}) display the results of
simulations for the two massive ion-ion collisions. The magnitudes
use  the calculated survival probabilities, normalised by the $pp$ or $p+D$
experiments.  The rather low $E_t$ value at which the measured J/$\psi$
suppression becomes pronounced obtains equally well in the simulation, and
the same low level of J/$\psi$'s is reproduced.  The results reinforce the
perception already created by the comparison with the minimum bias data. The
hadronic two-step cascade is capable of describing the charmonium yields,
J/$\psi$ and $\psi'$ as well. Again, the source of the suppression is
multifold, perhaps half from comover interaction and half from the hard
cascade, both strongly influenced by our treatment
of the charmonium states. The beginning of strong suppression in J/
close to peripheral collisions, $E_t \sim 50$ GeV, is a reflection of the
role the heavier charmonium states play, and  occurs at or near the
correct observational point.

\section{Conclusions}

It appears that a conventional hadronic explanation of the minimum bias
and central J/$\psi$ and $\psi'$ suppression in A+B collisions is
possible. This is accomplished here with a cascade, not specially tuned for
the charmonium sector alone, but consistent with soft energy loss processes,
both meson and proton spectra, and with a broad range of Drell-Yan
data. There are parameters in the model, notably the meson formation time,
the generic meson decay constant, and certainly the elements of the
charmonium reaction matrix. However, we have not made use of all of this
freedom in obtaining the main results. We have used a pure diagonal reaction
matrix to describe the coupled charmonium channels, no mixing was needed
further than that implied in $NN$. The ratio of charmonium states produced in
the elementary nucleon interactions is naturally somewhat uncertain, but
strong hints are available from experiment and we followed these indications.

Many of the other parameters are also constrained by p+A data or by
reasonable assumptions, for example the relative size of breakup
probabilities must be tied to the relative sizes of the charmonium states.
The formation time for produced mesons is also strongly tied to inclusive
meson production. Some authors \cite{Kharzeev} suggest theoretically that the
J/$\psi$ total cross-section on mesons must be drastically smaller than the
$4$--$6$ mb we use for breakup, but this argument is disputed by others
\cite{Huefner2}, and it seems unlikely. Our direct experimental knowledge of
the total and partial cross-sections, including any energy dependence, for
J/$\psi$ or other charmonium mesons on the lower mass mesons is of course
very limited.
 
Evidently, the initial phase of our model, or that of others, for charmonium
creation or breakup is not strictly nor conventionally hadronic. The
production and destruction of states in this phase cannot be of standard
fully formed charmonium states alone. But whether one imagines these initial
forms to be preresonant or a somewhat excited $c \bar c$ may not change the
effective dynamics. We have chosen to view the early life of the hidden charm
mesons as akin to a $c$ and $\bar c$ in relatively narrowly defined continuum
wave packets and characterised by their relative distance.  It is crucial
that the $c \bar c$ pair be in both colour singlet and octet states. Then,
despite the lengthier formation time for the more massive states, in fact
precisely because of this, these states are harder to form, suffering more
collisions. The dependence of an effective cross-section on radius then
follows.

For an elementary pp collision the eventual coalescence of
this pair into a proper charmonium state could be described in a fashion
following the formation of deuterons or other clusters \cite{clusters}.
In the case of $pp$ there would be independent formation probabilities for
the various spectroscopic states, but the reaction matrix would presumably
be diagonal, assuming no significant final state interactions on the charmonium.
The coalescence within a nuclear environment would be very different than in
free space, with many more interactions preventing bound state formation.
The more general matrix we proposed, $R_{ij}$, would have a simple meaning in
an extension of this picture to ion-ion collisions, its elements describing
what happens to these early $c \bar c$ states as they propagate through the
nuclear environment, above and beyond their fate in pure $pp$.

What then has been learned about excited, dense, nuclear matter from the
reduction in J/$\psi$'s? Our earlier calculations \cite{LUCIFERII} for a
broader range of processes, suggested that very high baryonic and mesonic
energy densities were achieved in central Pb+Pb interactions, $\rho_B\sim
6$--$7$/(fm)$^3$ and $\rho_E\sim 4$--$5$ GeV/(fm)$^3$ respectively and that
these densities persist for quite long times $\tau \sim 3$--$5$ fm/c, in the
CM frame. Thus appropriate conditions for possible `plasma' creation exist in
the most massive collision. This matter density has been sensed in the
theoretical comover breakup of charmonium, both J/$\psi$ and $\psi'$, more so
for Pb+Pb than for S+U. But should our model stand the test of time, and it
has explained a good portion of existing data at the SPS, then the case for a
non-conventional explanation is hard to establish as yet.  It is always of
course possible that partons actually are playing a less passive role than
portrayed by hadronic modeling, that especially high gluon densities are achieved in the
initial phase. If so, the use of purely hadrons seems to mock up
the partonic behaviour rather closely, and even the early charmonium and
Drell-Yan dynamics (consider, {\it e.g.\ }p+A) seem to, effectively at least,
fit into this picture. In any case our simulations do not rule out the
creation of some form of partonic matter in ion collisions at SPS
energies. They only make the necessity thereof less compelling. A
microscopic, non-hadronic treatment of the internal strusture of the
charmonium states might alter the entire picture, and may be necessary at
higher energies.

At RHIC energies and beyond, surely even the soft hadronic measurements will
be contaminated by contributions originating in harder parton interactions,
degraded say, by later hadronisation. Nevertheless, it may be hard to find
partonic footprints in the soft spectra. The generic mesons and baryons may simply
be stand ins for mini-jets \cite{eskola}, effectively and correctly
incorporating divided parton cross-sections and hadronisation within
hadronic collisions, decays and meson production. If appreciable partonic plasma is
quickly formed, in a state close to thermalisation, the loss of energy to the
baryonic background might very well be detectable. We intend to pursue such a
path to finding `signals' of collective parton dynamics. 

The dynamics of charmonium suppression would seem {\it ab initio} to be
similar at RHIC, but the greatly reduced duration of the initial hard cascade
at $s^{1/2}=200$ GeV might play a spoiler role. The quasi-hadronic
modeling employed in this phase by all practitioners may be inappropriate at
such increased energy.

\begin{ack}

The authors are grateful to J.~Huefner for stimulating and incisive comments,
and for bringing to our attention the at times exothermic behaviour of the
charmonium breakup cross-section near threshold. This manuscript has been
authored under US DOE grants NO. DE-FG02-93ER407688 and DE-AC02-76CH00016.

\end{ack}

\clearpage

\clearpage

\clearpage

\clearpage

\clearpage

\clearpage

\clearpage

\clearpage

\clearpage

\clearpage

\clearpage

\clearpage

\clearpage

\clearpage

\end{document}